\documentclass{article}
\usepackage{graphicx} 
\usepackage{soul}
\usepackage{authblk}
\usepackage{lscape}
\usepackage{xcolor}
\usepackage{color}

\title{Towards using renewable energy in Mezcal production}
\author[1,4,5]{Jesus Antonio del Río Portilla}
\author[1]{Argelia Balbuena Ortega}
\author[1]{Anabel López-Ortiz}
\author[1]{Jorge Alberto Tenorio}
\author[1]{Nicté Yasmín Luna Medina}
\author[1]{Patricio Javier Valadés Pelayo}
\author[2]{Federico del Río Portilla}
\author[2]{Mayra León-Santiago} 
\author[3,4]{Alfonso Valiente-Banuete} 

\affil[1]{Instituto de Energías Renovables,
Universidad Nacional Autónoma de México.}
\affil[2]{Instituto de Química,
Universidad Nacional Autónoma de México.}
\affil[3]{Instituto de Ecología,
Universidad Nacional Autónoma de México.}
\affil[4]{Centro de Ciencias de la Complejidad,
Universidad Nacional Autónoma de México.}
\affil[5]{Centro de Estudios Mexicanos UNAM Reino Unido, Universidad Nacional Autónoma de México.}
\date{\today}

\begin{document}

\maketitle

\begin{abstract}
 This paper explores the electrification of mezcal distilling in Oaxaca, Mexico, as a sustainable alternative to traditional firewood methods. We investigate the mezcal process, including cooking, grinding, fermentation, and distillation, and propose a photovoltaic system for distillation. The research also includes scientific outreach activities in the producing communities. We, in collaboration with the communities, proposed novel uses of renewable energies. The results of chemical analysis (chromatography and FTIR) and sensory data for distillation using firewood and electricity are presented to compare the mezcal produced with solar energy and traditional mezcal. Our studies conclude that electrical distillation can reduce environmental impact and improve energy efficiency without compromising product quality.
\end{abstract}

\section{Introduction}

Renewable energy holds significant relevance across remote, rural communities in Latin America. Specifically, to enhance economic development, improve the quality of life, and promote sustainable practices in remote communities. Among the economic impacts, we can mention that renewable energy facilitates access to electricity and modern tools and technologies,  enhancing agricultural productivity. With reliable electricity, farmers can utilise electric-powered equipment, improving their output and reducing labour costs and use of fossil fuels or pressure the forest to harvest wood for fire \cite{moreira}. Also, access to electricity can stimulate local economies by enabling small businesses to operate more effectively. Thus, access to electricity can enable new enterprises in sectors such as food and beverage processing, textiles, and services, creating jobs and fostering economic diversification in rural areas. The relevance of electrification in rural production in Mexico, as in any other Latin American country, extends beyond mere access to electricity; it encompasses economic growth, social improvement, and environmental sustainability, making it a critical component of rural development strategies \cite{moreira}.
According to the United Nations COMTRADE international trade database, Mexico's exports of beverages, spirits, and vinegar totaled US\$11.65 billion in 2022 \cite{tradingeconomy}. The top three Mexican-exported beverages are beer, tequila, and mezcal; while large companies produce the first two, small distillers in rural areas of Mexico produce mezcal. Mezcal is a Mexican spirit that has experienced exponential demand growth, underscoring the need to ensure sustainable and equitable production for both society and ecosystems. Mezcal production represents an opportunity to combine tradition and technological innovation, biodiversity conservation, economic development, and social welfare. Scientific research on biological, ecological, social, economic, and engineering topics is essential to ensure sustainable and equitable production, which benefits both producing communities, consumers, and ecosystems. The information available reveals the complexity of agroecological systems associated with agave cultivation.

Before proceeding with this study, we summarise the mezcal production process according to the Official Mexican Standard and the consensus of mezcal producers.

\subsection{Mezcal }

Mezcal is a rich tapestry of Mexican tradition, craftsmanship, and natural physical-chemical methods, deeply rooted in Mexico's cultural heritage. Let us outline the general steps for transforming the agave plant into a complex spirit known as mezcal. The fabrication begins in Mexico's arid landscapes, where the agave plant thrives under the relentless sun. It takes several years, sometimes decades, for the agave to reach maturity. When the time is right, skilled  \textit{jimadores}, or agave harvesters, carefully remove the leaves, revealing the heart of the plant, known as the \textit{piña}. Many agave plants are used to produce mezcal; their piñas are rich in sugars and flavours. The piñas are then transported to the \textit{palenque}, the traditional mezcal distillery, where they are cooked in earthen holes lined with rocks and covered with agave leaves, soil, or other secrets of the \textit{Maestros Mezcaleros}. These specialised artisans play a crucial role in the process, and they transmit their knowledge and techniques from generation to generation. 
The cooking process can take several days, during which the piñas are transformed, their sugars caramelising, and their flavours and aromas deepen with the Mezcaleros' secrets. Once cooked, the piñas are crushed to extract their sweet juices. Traditionally, this is done using wooden clubs (\textit{marros}) or a \ textit{tahona}, a large stone wheel pulled by a horse or mule; however, mechanical mills are also used in some mezcal variations. The crushed agave is then placed in large vats, where it ferments naturally. These vats can be made with wood, cowhide, or clay. The wild or specifically selected yeast interacts with the sugars in the agave, converting them into alcohol. This fermentation process can take up to two weeks,  during which the mixture, known as \textit{mosto}, develops a complex array of flavours and aromas. The final step in the mezcal process is distillation in copper or clay stills, which are heated to separate the alcohol from water and other impurities. The first distillation produces a low-alcohol spirit known as \textit{ordinario} or \ textit {llano}. This first product is then distilled a second time to refine the spirit, resulting in the final mezcal. Mezcal distillation is a delicate balance of art and science, requiring the skill and expertise of the Maestro Mezcalero to ensure the spirit retains its unique character.

The Official Mexican Standard NOM-070-SCFI-2016 outlines the specifications for the production, packaging, and marketing of mezcal, including its various categories (mezcal, craft, ancestral) and classes (young, rested, old, etc.). It defines the physicochemical characteristics it must meet and sets the labelling requirements for the domestic and international markets. Processes of processing, from raw material to packaging, are described. The mezcal should be produced exclusively from agaves grown in the geographical area delimited by the Resolution granting protection to the designation of origin. According to the elaboration process, the rule establishes three mezcal categories: Mezcal category is characterised by the use of modern methods and equipment in its elaboration, Mezcal Artesanal (Craft mezcal involves traditional techniques and tools in its production process), and Mezcal Ancestral is distinguished by using older and more rustic methods and equipment in its elaboration. Each category has specific maguey cooking, grinding, fermentation, and distillation requirements as shown in Table \ref{Tcategory}.

\begin{landscape}
\begin{table}[]
    \centering
    \begin{tabular}{|c|c|c|c|c|}
    \hline
     Category & Cooking & Grinding & Fermentation & Distillation  \\
     \hline
     \hline
     Mezcal    & well ovens&  heartbreaking & wooden containers  & still \\
                & clay &  Chilean or Egyptian mill & masonry pools& continuous distillers \\
                & autoclave & mill train  & stainless steel tanks & of copper or stainless steel.\\
                \hline
    Craft Mezcal & well ovens &  ``tahona'' & holes in stone, soil or trunk & \textbf{Direct fire} in copper \\
     & masonry highs & Chilean or Egyptian mill & masonry pools & or stainless steel stills\\
     & & heartbreaking & wood or clay containers & \\
     \hline
     Ancestral Mezcal & well ovens & ``tahona'' & clay container & \textbf{Direct fire} \\
      & & or Chilean or & holes in stone &in clay containers \\
            & & Egyptian mill & wood, clay containers &\\
             & & & animal skins containers & \\
    \hline
    \end{tabular}
    \caption{Characteristics of the different mezcal categories according to the Mexican Standard appeal of origin}
    \label{Tcategory}
\end{table}
\end{landscape}

We have emphasised the specific requirement of \textbf{direct fire} in the distilling process. It is worth stressing that four other Mexican spirits with designation of origin do not include the \textbf{direct fire} distilling process in their Official Mexican Standards. After analysing the distilling process for mezcal, direct fire appears to play little or no role in imparting mezcal's special features. Moreover, using wood to distil is causing severe damage to the natural forests of the mezcal-producing regions. Therefore, the electrification of distilling mezcal deserves a community participation test, isolating and assessing how replacing direct firing during the distillation stages influences the characteristics of the final product.

On the other hand, the electrification of dairy activities in Mexican rural areas has been around for several decades, revolving in the renewable energy sector \cite{manuel88}.
The electrification of rural production in Mexico is highly relevant for enhancing economic development, improving the quality of life, and promoting sustainable practices in remote communities \cite{moreira}. Electrification facilitates access to modern tools and technologies, enhancing agricultural productivity and efficiency in rural areas. Farmers can use reliable electricity-powered equipment, improving their output and reducing labour costs. Access to electricity can stimulate local economies by enabling small businesses to operate more efficiently and creating opportunities for new enterprises in food processing, textiles, and services. However, the impact of electrification goes beyond economics; it transforms how people live and work, enhances their quality of life, and opens new economic and social opportunities. The shift to renewable energy is relevant for electrifying daily activities because it minimizes environmental impact compared to traditional fossil fuels, reduces greenhouse gas emissions, and promotes sustainable development practices in rural areas. 
The relevance of electrification to rural production in Mexico extends beyond mere access to electricity; it encompasses economic growth, social improvement, and environmental sustainability.

However, we stress that before shifting to the electrification of rural mezcal production, it is necessary to demonstrate that mezcal distilled with electric energy and through ancestral distillation methods are equivalent. Therefore, this work uses a sensory evaluation of alcoholic beverages involving producers. 
Moreover, this paper describes a socio-technical exploration of mezcal production, evaluating the feasibility of electrifying the mezcal distilling process to avoid firewood use and reduce current environmental pressures, while retaining the social and organoleptic advantages of the traditional distilling process.

We organise the rest of the paper as follows: We present the methodologies used to exchange knowledge with two mezcal producer communities in Oaxaca, Mexico's highest mezcal producing region, in section \ref{exchange}. In these session discussions, we find exciting energy requirements beyond the traditional energy used for cooking and distilling parts. In section \ref{electric}, we describe the electric distillation procedure. Section \ref{samples} describes mezcal samples and reagents we use in our study. Section \ref{sensoryP} details the alcohol, physical, and chemical quantification and sensory evaluation methods. It is important to emphasise that the sensory evaluation was performed double-masked among mezcal producers. Our main findings are presented in section \ref{results}, where sensory, chromatographic and FTIR spectra are presented and analysed. 
Finally, we conclude the paper with lessons learned and valuable comments for future activities.

\section{Knowledge exchange activities}\label{exchange}

In 2022, we visited palenques to engage with mezcal producers and exchange knowledge. We used a photovoltaic prototype to generate electricity and distil a mixture of water and ethanol, thereby separating alcohol from water. Two science communication activities were conducted using a democratic science communication model \cite{alcibar}. The objectives were to build trust within the community of Santo Domingo Yanhuitlán, Oaxaca, as well as among mezcal producers in the region, and to share the energy-related challenges and potential solutions in the mezcal production. Science communication can serve as a tool that enables communities to value knowledge in line with their own interests and adopt technological solutions \cite{olive}. The democratic science communication model argues that the dialogue between scientists and society is necessary to reach consensus on technological solutions.\cite{alcibar}.

We performed a puppet play in the Santo Domingo Yanhuitlán, Oaxaca community, dedicated mainly to children but open to the general public. This play recounts the battle between fossil and renewable energy sources in a Mexican setting (Mexican wrestling). Also, we held workshops with mezcal producers to exchange knowledge about mezcal fabrication and diverse energy uses. In particular, we presented the photovoltaic prototype in action under sunlight in an open space, separating water and alcohol. After that, we had meetings, dividing the producers into teams to discuss energy use in the mezcal process. Of course, one of the main ideas was the option to use solar or electric energy in the last part of mezcal production, distillation. These last activities were performed in different communities in Oaxaca State. 

\subsection{Discussion sessions with mezcal producers}

As we mentioned, the mezcal producers raised questions about using renewable energy, but most surprisingly, they identified new energy requirements in different parts that we, scientists, had not noticed. 

\begin{itemize}
  \item Energy for fermentation: Maintaining stable fermentation temperatures is essential, especially during winter. Low temperatures are common in the high mountainous regions, so heating is required to maintain a year-round steady production. After the discussions, we proposed exploring the possibility of using renewable energy, such as solar heating panels, to regulate the temperature of the water needed for fermentation.

  \item Cooling water in distillation: Besides heating, the distillation process requires large amounts of cold water to condense the alcohol vapour. Producers worry about the disposal of hot water, and many recycle it through a closed system. However, the efficiency decreases as the temperature of the water condensation outflow increases. The distilling stage takes more than 48 hours of continuous work. Thus, using solar heaters at night to cool water is feasible (i.e., via radiative heat exchange with the sky).
  
\end{itemize}

In these conversations, we opened our minds. We considered using standard solar heaters during the day to heat the fermentation process and during the night to cool the water from the condensation stage. These are the main results of the knowledge exchange.
These discussions demonstrated a strong interest among producers in improving energy efficiency without sacrificing mezcal quality, opening the door to innovations that maintain traditions while integrating more sustainable practices.

Finally, we all agreed for now to use a prototype system consisting of two photovoltaic panels, an inverter, and a battery to store electric energy. The prototype is intended to replace firewood for distillation while being more affordable than gas. We plan a second working visit to palenques to distil the \textit{mosto} with them, but before proceeding with the evaluation, we present some interesting details about the fabrication procedures across different communities.

\section{Mezcal fabrication procedure}

In this section, we go beyond the general procedure and describe in detail the qualitative and quantitative aspects of mezcal fabrication in Oaxaca's region. Mezcal preparation involves several stages: harvesting (or selecting the agave), cooking the agave heart (a.k.a. piña), grinding, fermentation, and distillation.

The oven design and materials are crucial for the agave cooking process. Several oven types can produce mezcal, including stone ovens, masonry ovens, and modern industrial steam ovens. All the ovens used by the communities we visited are of the first type, which is the most traditional. These are built by digging a hole in the ground, covering it with stones, and using soil as a sustainable source of thermal mass and insulation. 


After harvesting, Medium-Sized, oblong-shaped stones are selected and preheated over a fire of firewood. The rocks range in size from 10 to 30 cm, large enough to retain heat for a few days while remaining amenable to manipulation. Stone preheating is typically done in the cooking oven, although it can be done elsewhere, such as in a fire pit, and the stones can be brought over later. 

This choice depends on whether the oven design is versatile enough to operate under different ventilation conditions. An ample air supply is required during rock preheating for efficient wood combustion. Reducing it during cooking helps the rock bed retain its heat for extended periods while encouraging the regulation of an equilibrium between combustion and agave's thermal hydrolysis and Maillard reactions.

During cooking, the agave piñas are placed in the oven with additional firewood, covered with preheated stones and leaves to allow for slow cooking. Temperature control, air supply, and the rocks-to-wood-to-agave ratio are essential to balance the flavours that will ultimately emerge during fermentation and distillation, giving mezcal a characteristic smoky flavour that is not perceptible with either masonry or industrial ovens. Since the characteristic flavour of mezcal is obtained at this stage, a renewable alternative is not currently conceivable. However, studying how these ovens work and their thermal characteristics can lead to research aimed at improving their efficiency.

Mezcal fermentation can be done in various containers, each with its particularities. The containers we found in the communities we visited were wood, clay, and cowhide. Wooden basins interact with the environment, allowing a greater influence of natural yeasts during fermentation, resulting in mezcals with more complex, artisanal flavours. Clay containers are used in more traditional processes. Clay allows for slow fermentation, which can enrich the mezcal's flavours. Cowhide is an ancestral container that gives mezcal a unique profile with animal and earthy notes.
During the winter, lower temperatures can slow down fermentation, affecting the process. To counteract this, producers must adjust the fermentation time or use heating to maintain the optimal temperature.
Distillation occurs in stills made of three different materials: copper, clay, and stainless steel. In this part, the types of mezcal produced are defined as traditional, ancestral, and industrial. Each of these definitions is associated with the kind of distiller used: traditional mezcal-copper, ancestral mezcal-clay, and industrial mezcal-stainless steel.
The activities we identified as more demanding (energetically speaking) are cooking the piña and distillation. Although there are no exact data or precise estimates on the consumption of firewood throughout the history of the mezcal industry, nor the consequent degradation of forest resources due to this activity, it is relevant to highlight that to produce one litre of mezcal, approximately 10 kg of firewood are required. In other words, with one $m^3$ of wood, 125 litres of mezcal can be obtained. Producers told us that only a third or a half of this wood is used to cook the piña, while the rest is used for distillation. Moreover, the use of wood in the distillation process has implications beyond its ecological impact, as in some  palenques (where mezcal is distilled), there is no ventilation, and the smoke can be inhaled by producers, compromising their health. Thus, based on this information, we proceed to explain the electric distillation stages.

\section{Electric distilling procedure} \label{electric}

Approximately 4 litres of fermented liquid are placed into a stainless steel still to carry out the distillation. This traditional distillation apparatus, known for its reliability, separates components of a liquid mixture by exploiting differences in their boiling points. The process involves heating the fermented mash in a boiler until specific components evaporate at their respective boiling points. For instance, under standard conditions, ethanol, the alcohol in spirits, evaporates at $78.4^\circ$C, which is lower than water's. The vapours then rise through a pipe into a condenser, where they cool and return to liquid form. This condensed liquid, the distillate, contains the more volatile compounds from the original mixture, often with higher alcohol content or purified essential oils. By controlling the temperature, we ensure that only the desired components evaporate and condense. In some processes, like the production of whiskey or vodka, multiple distillations are performed to improve purity and flavour.

The still we used is composed of a $9.8'' \times 9.8''$ $ (25 \times 25 cm)$ stainless steel pressure-sealed pot and a condenser $7.9'' \times 4.3''$ $(20 \times 11 cm)$.
They still adopt an open cooling method. The high thermal conductivity of the copper coil allows it to reach $73^\circ$ through heat transfer. However, the cooling system must be sufficiently efficient due to this thermal conductivity.

Our still is equipped with a real-time thermometer on the lid, featuring a dual display of Celsius and Fahrenheit for convenient temperature monitoring inside the pot during distillation. We also used a thermographic camera to follow the temperature of the pot and the copper coil. We used food-grade silicone tubes, which are flexible and robust to ensure a good seal. Four buckles and silicone gaskets within the lid provide a tight seal.


\section{Samples}\label{samples}
A total of three samples were analysed, labelled as follows: mezcal from a solar distillation carried on in Santo Domingo Yanhuitlán, Oaxaca (Solar), mezcal from a solar distillation carried on at IER-UNAM (Solar 2), mezcal from an ancestral distillation without rectification (llano), and mezcal from an ancestral distillation with rectification (ancestral). 
The samples were collected directly from the distillation process and stored at room temperature until analysed.

Absolute methanol (Merck 99.5\%) and absolute ethanol (Sigma Aldrich 99.5\%) were used as comparison standards. To compare with a commercial traditional mezcal, we bought a bottle in the region of San Juan del Río, Oaxaca.

\section{Physical, Chemical and Sensory characterization} \label{sensoryP}

This section presents the sensory evaluation, FTIR, and gas chromatography methodologies used in our analysis.

\subsection{Double blind trial on sensory features}

\begin{figure}
    \centering
    \includegraphics[width=1\linewidth]{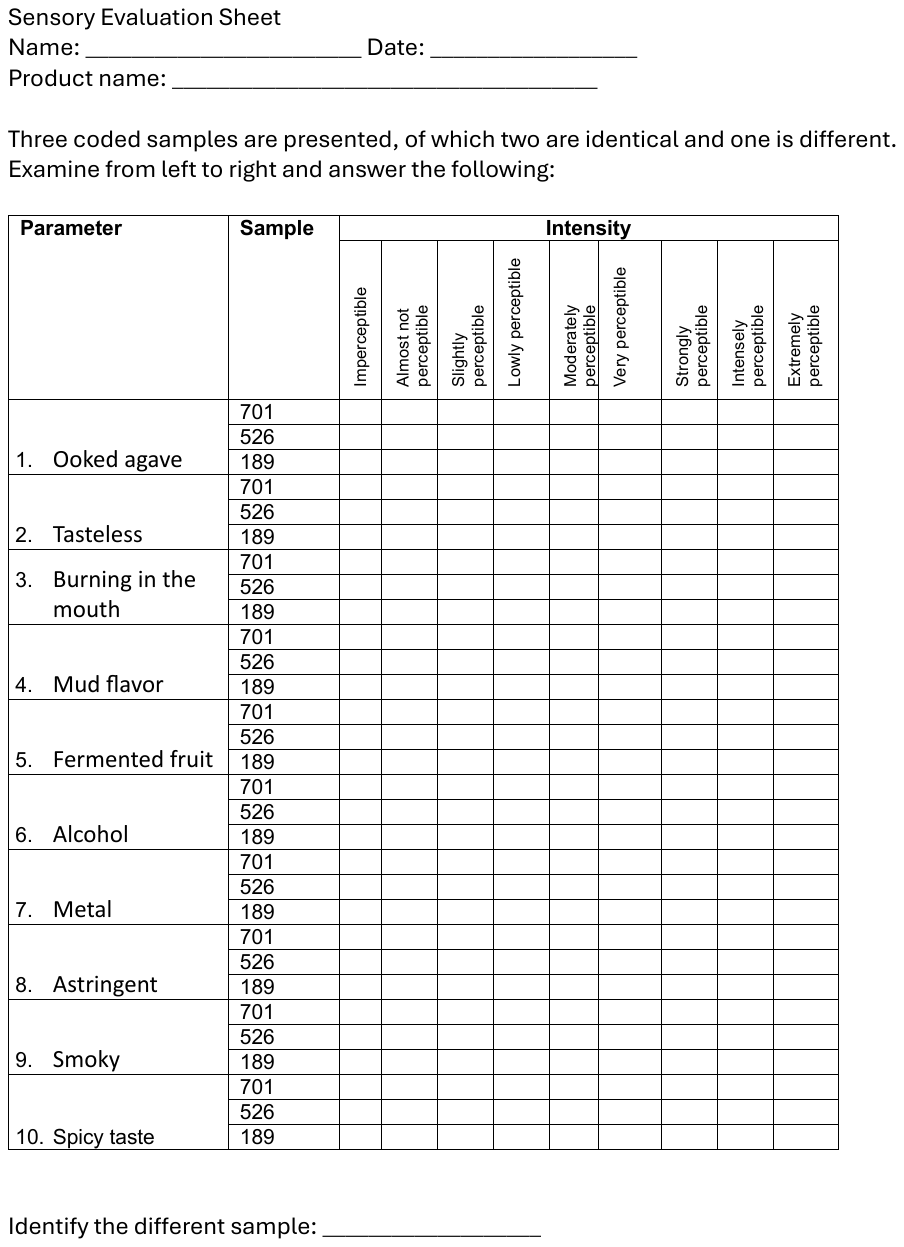}
    \caption{Sensory test applied to Maestros mezcaleros}
    \label{sensory}
\end{figure}

The sensory characterisation was carried out using a hedonic scale of 9 points. The parameters selected to perform the intensity evaluation were chosen from the attributes mentioned by Barajas et al. \cite{BARAJASRAMIREZ2024}. We selected 10 attributes: cooked agave, tastelessness, burning in the mouth, mud flavour, fermented fruit, alcohol, metal, astringent, smoky, and spicy taste on an intensity scale from zero to nine. The sensory attributes were evaluated according to Table \ref{sensory}. The total number of non-trained participants was 24 (n=24). All participants were maestros mezcaleros from Oaxaca, Mexico. 

Principal components analysis (PCA) was performed using all attributes used in the intensity analysis. The matrix was adjusted with the variables and the number of observations. The principal components were obtained by subtracting the mean from each observation. The matrix obtained from the principal components was analyzed using NCSS 2020. The correlation between the variables was calculated using a Pearson correlation method using the following equation \ref{pearson}.

\begin{equation}
    r=\frac{\sum_{i=1}^{n}(X_i-\bar{X})(Y_i-\bar{Y})}{\sqrt{\sum_{i=1}^{n}(X_i-\bar{X})^2\sum_{i=1}^{n}(Y_i-\bar{Y})}}. \label{pearson}
\end{equation}
Where values near 1 show a strong correlation and values near zero show a weak correlation. The analysis was performed using NCSS 2020 software.

\subsection{FTIR evaluation}

IR spectroscopy has been established as a tool for quality control in different industrial processes \cite{Schulz2007}. ATR-FTIR spectroscopy has been used to identify the content of organic compounds in mezcal \cite{Lopez-rosas}. 

A Thermo Fisher spectrometer equipped with a diamond attenuated total reflectance device was used to analyze the mezcal samples. A total of 50 scans were made with a resolution of 4 $cm^-1$. The gain was 1 $cm^-1$ with an optical speed of 0.4747 at an aperture of 100\%. The background measurement was made before the analysis. The ATR crystal was cleaned before each sample measurement with distilled water and ethanol added to  paper tissue. Ethanol and methanol were used as standards to determine the differences between the samples and the standards. The samples, ethanol, and methanol were diluted to 50\% with water. The measurements were made in triplicate. 

\subsection{Methanol quantification}
To determine the methanol content in the four samples, we used 10 ml and added 1-Hexanol (1-H) as an internal standard, such that the amount of sample used had a concentration of 97.68 mg (hexane) / 100 ml of distilled solution. The sample was then injected into the gas chromatography system with a flame ionization detector (CG-FID 7890B from Agilent Technologies) using the Open Lab acquisition software.
Chromatographic analysis was performed using a DB 624 capillary column, 30 m long, 250 $\mu m$ in diameter and 1.4 $\mu m$ in thickness, with an injection volume of 1 $\mu L$ and a split ratio of 30:1.
The temperature conditions were: injector: 250 °C, detector: 250 °C, program: 40°C for 1 minute, ramp 1 at 8 °C/min, 120 °C for 1 minute, ramp 2 at 15 °C / min, and 240 °C for 2 minutes. The mobile phase used was helium at a flow rate of 1 mL/min

\section{Electric and traditional distillings are statistically indistinguishable} \label{results}

We organise the results as follows: first, we present the sensory evaluation, which shows no significant differences. Also, the results of the FTIR analysis of distilling samples and standard alcohol reagents present similarities. Finally, this section presents an HPLC analysis that corroborates the previous results.

\subsection{Sensory evaluation}
\begin{figure}[h]
    \centering
    \includegraphics[width=1\linewidth]{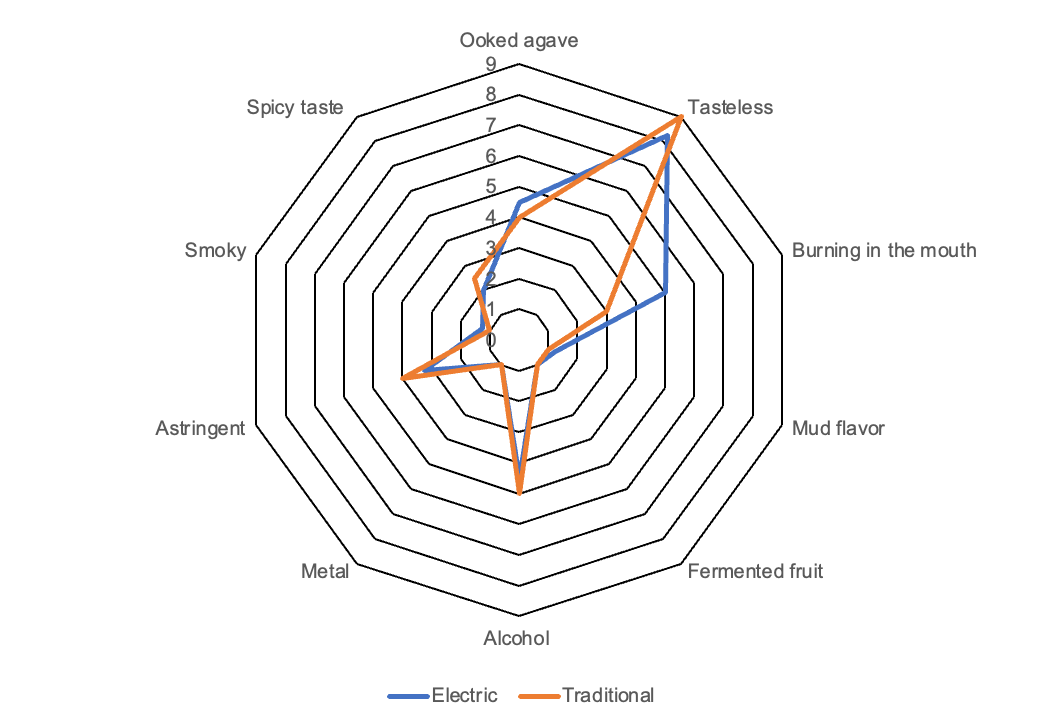}
    \caption{Sensory analysis of mezcal samples from solar distillation (electric), ancestral distillation (ancestral).}
    \label{Sensorial}
\end{figure}

The sensory evaluation results are shown in Figure \ref{Sensorial}. The profile of the samples corresponds to that of the solar mezcal (electric) and that of the ancestral mezcal (ancestral). The data had a normal distribution and homoscedasticity of variances. It was determined that there is no significant difference between the mezcal obtained by ancestral distillation and the mezcal obtained by electric distillation. All evaluated parameters: cooked agave, tasteless, burning in the mouth, mud flavor, fermented fruit, alcohol, metal, astringent, smoky, and spicy taste have a homoscedasticity of the medians.

\begin{figure}
    \centering
    \includegraphics[width=0.75\linewidth]{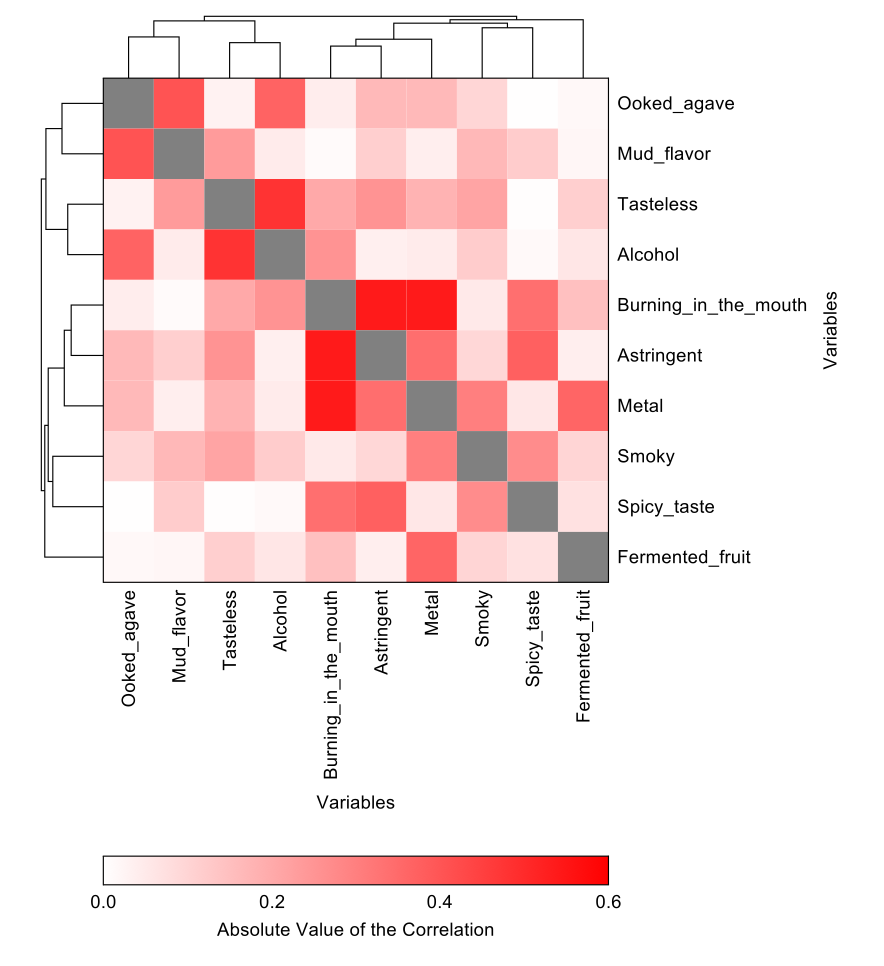}
    \caption{Headmap of the absolute values of the correlation matrix}
    \label{head}
\end{figure}

The heatmap of the absolute values of the correlation matrix allows us to visualize the strength of relationships among variables in mezcal sensory attributes (Figure \ref{head}). The stronger correlation occurs among the metal, astringent, and burning-in-the-mouth parameters. The value of correlation is 0.45, 0.48, and 0.7  respectively.

\begin{figure}
    \centering  \includegraphics[width=0.8\linewidth]{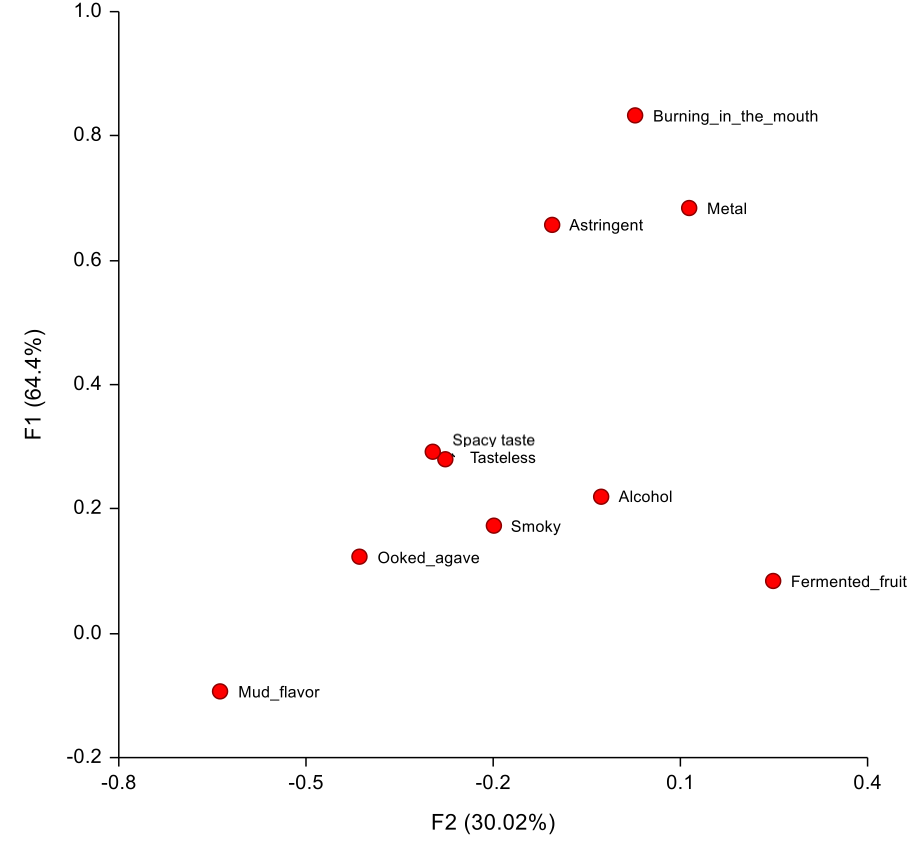}
    \caption{Principal factor analysis of the evaluated attributes of mezcal}
    \label{PFA}
\end{figure}

The second set of correlated variables is cooked agave and mud flavor, with 0.41 and 0.17 values, respectively. Figure \ref{PFA} shows the relationship between the sensory attributes evaluated in this work (alcohol, fermented fruit, burning in the mouth, etc.). Two principal factors derived from factor analysis, labeled as F1 and F2 were analyzed. F1 and F2, are  64.4\% and 30.02\% of the variance, respectively, showing they essentially capture the sensory data structure. 

Burning in the mouth, Astringent, and Metallic show significant positive loadings on F1, suggesting F1 represents a factor linked to strong or aggressive mouthfeel. On the other hand, Mud flavor and Cooked agave have negative loadings on F2 and moderate ones on F1,  indicating that F2 aligns with less desirable earthy or smoky attributes. Fermented fruit and Alcohol, positioned to the right (high loading on F2), might associate F2 with fermentation and alcohol-related traits. Attributes like Tasteless, Spicy taste, and Smoky, located near the center, less distinctly influence the factors, implying they are more balanced or undefined within the sample set.

\subsection{FTIR analysis}

\begin{figure}[h]
    \centering
    \includegraphics[width=0.8\linewidth]{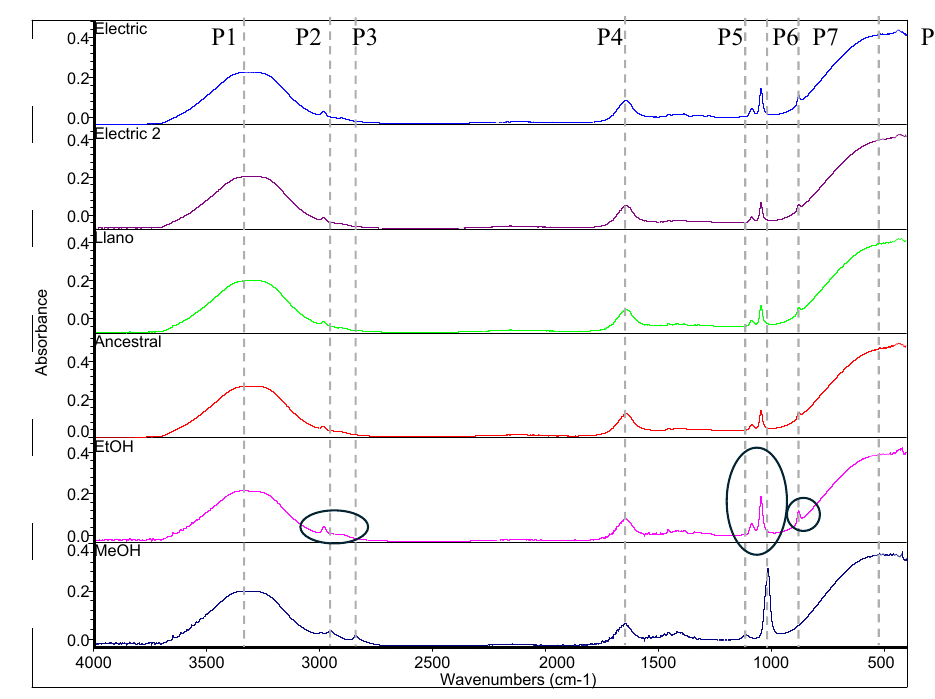}
    \caption{FT-IR analysis of mezcal samples from solar distillation in Santa María Tequisistlán (electric 1), solar distillation in Santo Domingo Yanhuitlán (electric 2)  ancestral distillation (ancestral), from the first distillation without rectification from ancestral processing (llano), methanol as reference (MetOH) and ethanol as reference (EtOH).}
    
    \label{FTIR}
\end{figure}

We analyze several samples of mezcal using FTIR spectophotometer. The results of the FTIR analysis are shown in Figure \ref{FTIR} main bands were identified in the Solar, Traditional, plain and ethanol samples: $P_1= 3330 cm^{-1}$, $P_2= 2980 cm^{-1}$, $P_3= 2839  cm^{-1}$, $P_4=1646  cm^{-1}$, $P_5 = 1085  cm^{-1}$, $P_6= 1044  cm^{-1}$, $P_7 = 876  cm^{-1}$, and $P_8=418  cm^{-1}$. Only 7 main bands were identified in the methanol sample: $P_1= 3330  cm^{-1}$, $P_2= 2950  cm^{-1}$, $P_3= 2839  cm^{-1}$, $P_4=1646  cm^{-1}$, $P_5 = 1112  cm^{-1}$, $P_6= 1013  cm^{-1}$ and $P_8=418  cm^{-1}$.

The peak $P_1$ has been attributed to the OH alcohol groups and is present in all samples. Using the standards included in MeOH and EtOH, we can identify that $P_1$ is present in both samples, which indicates O-H stretching due to the presence of alcohols, water, and hydrogen bonding. This indicates that all samples contain hydroxyl groups and water. The peak $P_2$ corresponds to the stretching of the C-H link, which is present in all samples associated with CH$_2$ and CH$_3$ groups. The intensity of this peak is lower in ancestral and electric samples
than in alcohol references. However, the $P_2$ peak is displaced to the right in the methanol sample. The $P_3$ corresponds to the vibration of stretching and deformation of the CHO link. This peak is also observed in all samples; however, the peak $P_3$ again shows a displacement in the methanol samples. The $P_4$ and $P_5$ peaks correspond to the elongation vibrations of the C=O and C-O links. $P_4$  indicates a clear presence in all ancestral and electric samples, indicating contributions from hydrogen-bonded water and oxygenated organic compounds such as organic acids or aldehydes. The mezcal and references (MeOH and EtOH) are a combination of alcohol and water; thus, the peak $P_4$ represents the bending mode frequency of the H-O-H\cite{Seki2020}.
In the case of methanol peaks, the displacement to the left of the peak $P_5$. The peak $P_6$ is attributed to the vibrations of the HC links and C-O stretching, suggesting the presence of alcohol bonds (C-O). The peaks are more visible in EtOH and MeOH than in mezcal samples. In methanol, a shift to the right of the P6 peak is observed. The peak $P_7$ is not present only in the ATR-FTIR spectrum of methanol. In addition, the ethanol spectrum coincides with the peaks observed in distilled solar mezcal, traditional mezcal, and Llano. Therefore, it is clear that the mezcal distillates specifically contain ethanol.

\subsection{Methanol content}

\begin{figure}
    \centering
    A)\includegraphics[width=0.45\textwidth]{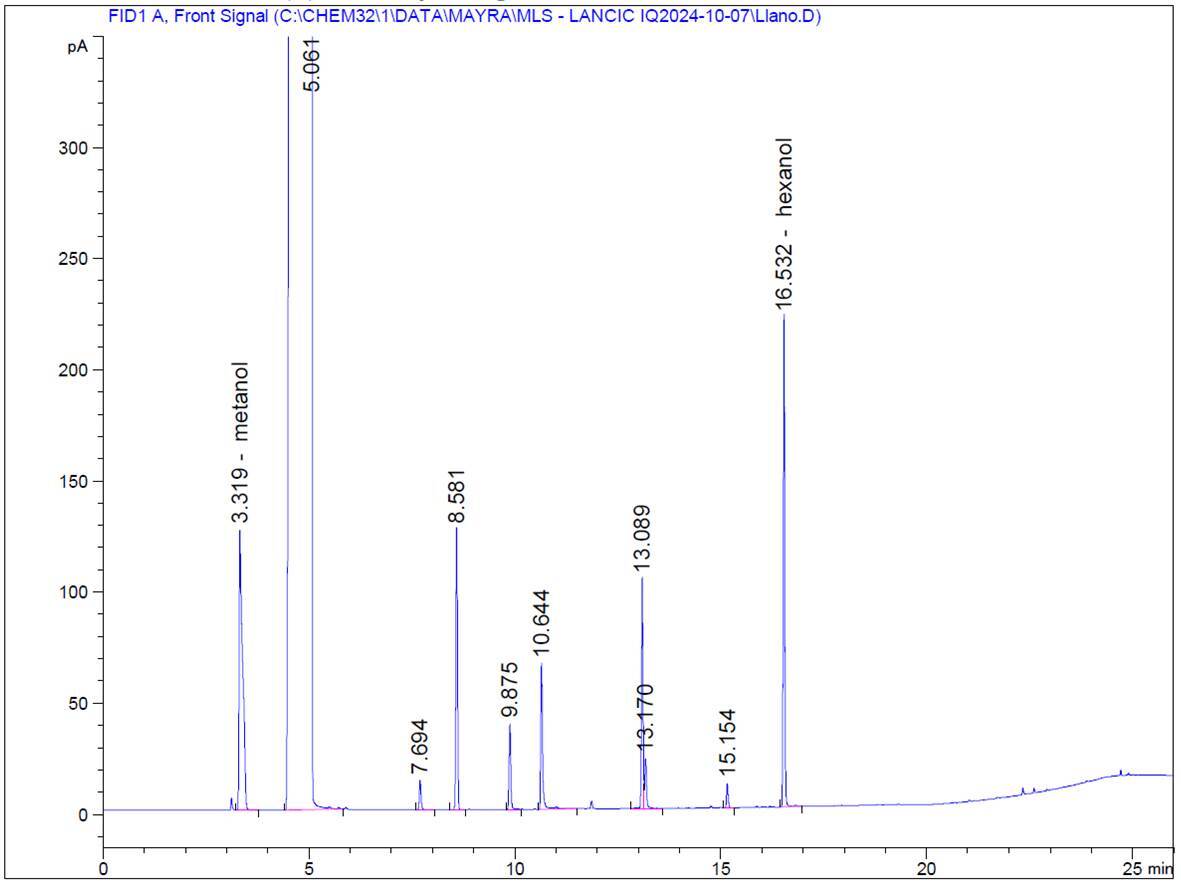}
    B)\includegraphics[width=0.45\textwidth]{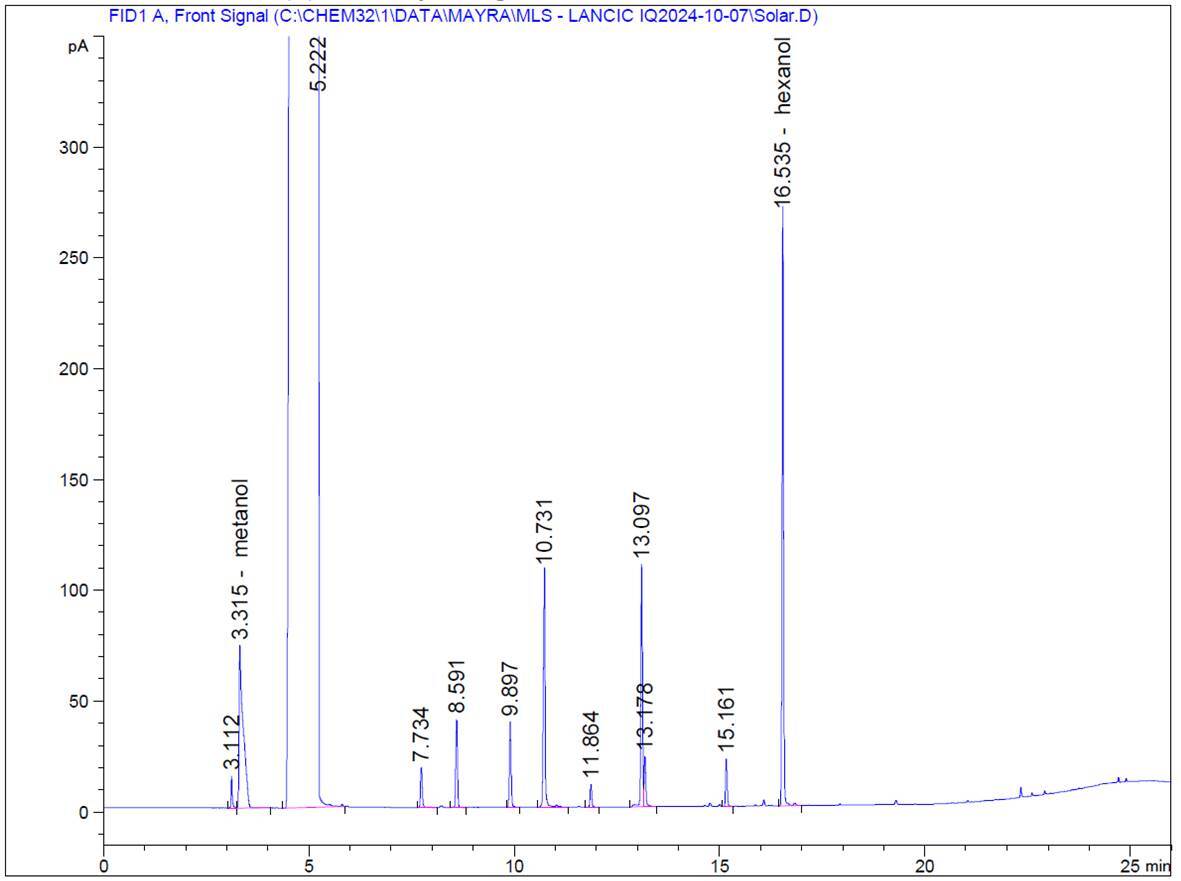}\\
    C)\includegraphics[width=0.45\textwidth]{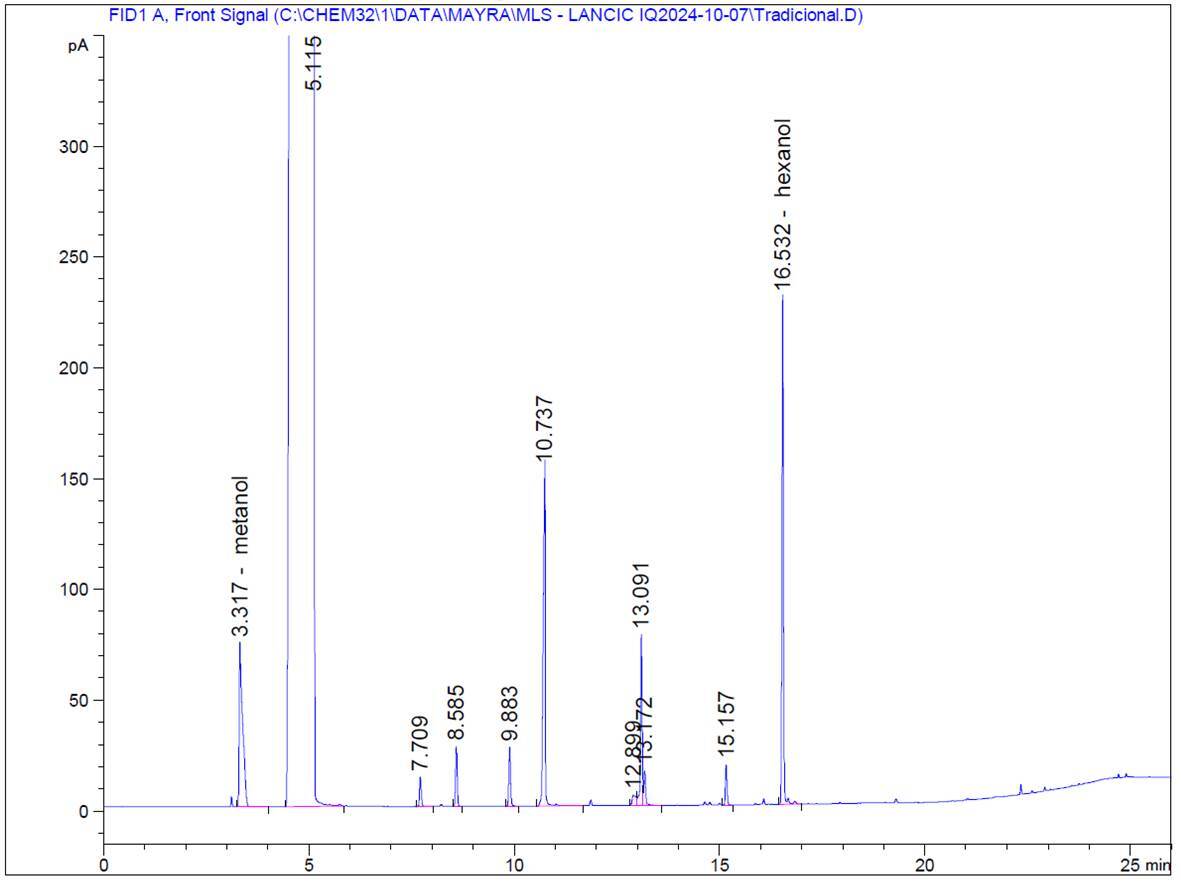}
    D)\includegraphics[width=0.45\textwidth]{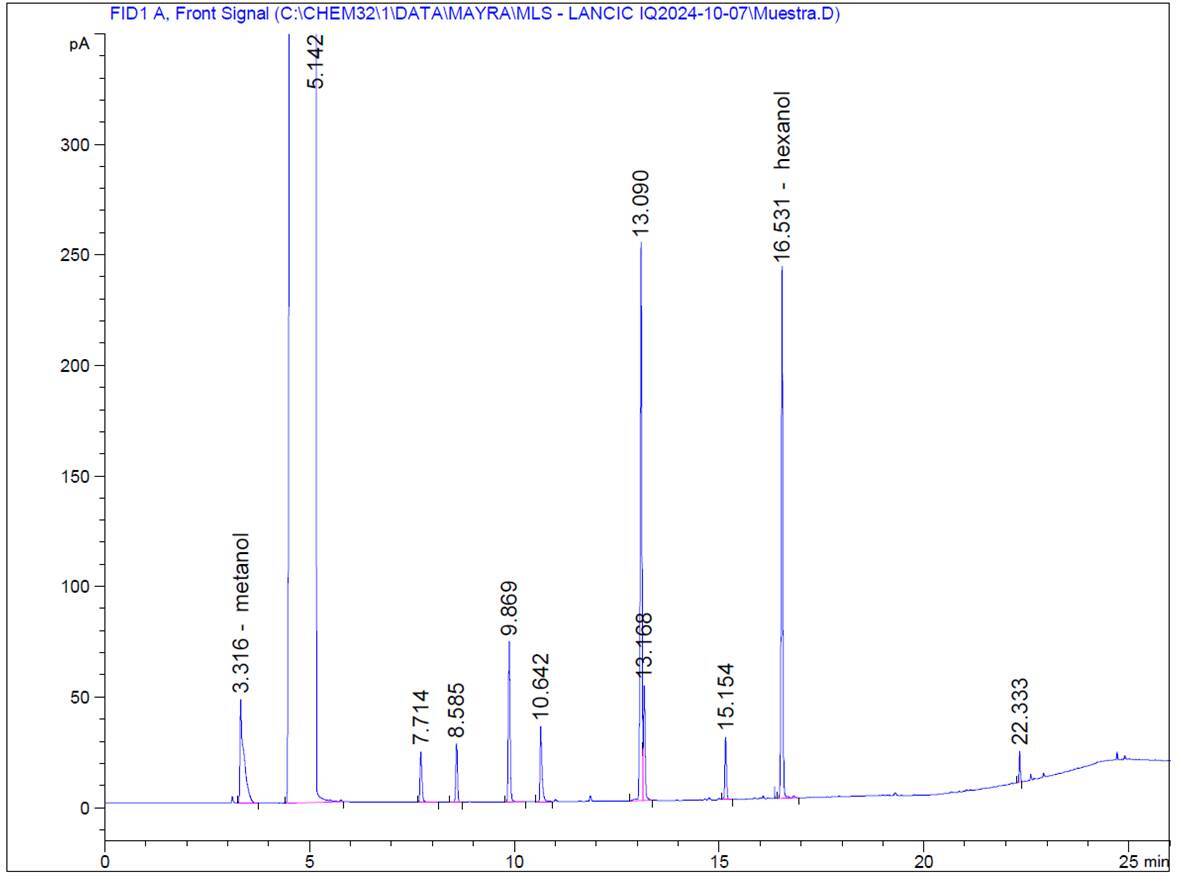}
    \caption{Chromatograph spectra of four mezcal samples: A)First electric distillation in a steel recipient (llano), B)second electric distillation in a steel recipient (electric), C) second fire distillation in a cooper recipient (traditional) and D) second fire distillation in a clay recipient (ancestral).}
    \label{gases}
\end{figure}

In this analysis, we used four samples. Methanol retention time is identified at minute 3.31; 1-Hexanol retention time at minute 16.53. The chromatogram of the samples shows the signals of methanol and 1-hexanol are in Figures \ref{gases}. Based on the areas of the chromatogram, the area ratios of Methanol/1-Hexanol for each sample were interpolated on a calibration curve. The results obtained for each sample are shown in Table \ref{tablegases}.

In Fig. \ref{gases} of samples A, C, and D, the same compounds were identified based on retention times of 7.6, 8.9, 9.9, 10.0, 13.1, 13.2, 15.2, and 16.5 minutes. All were identified using the same retention time. Sample B showed an additional compound at 11.9 minutes in very low quantities (see Figure \ref{gases}C). Despite this additional compound, it can be concluded that there are no significant differences between the four samples, except for the methanol content detected. The differences lie in the methanol content determined for the four samples (see Table \ref{tablegases}). Sample A contains 273.68 mg/100 mL of distillate. Samples B and C have almost identical methanol concentrations, while Sample D has the lowest methanol content.

\begin{table}[]
    \centering
    \begin{tabular}{|l|r|}
    \hline
    Sample & $mg_{methanol}/100 ml$\\
    \hline
    \hline
     llano    & 273.68  \\
     electric & 152.67 \\
     traditional    & 150.94\\
     ancestral & 98.05 \\
     \hline
    \end{tabular}
    \caption{Showing the quantitative methanol results for the representative samples}
    \label{tablegases}
\end{table}

Therefore, from the results of the sensory evaluation, FTIR, and gas chromatography, we conclude that the electrical distillation of the agave is an alternative to replacing the use of firewood. Moreover, implementing this alternative should help reduce the amount of firewood used in distillation by about 50\% to 67\% of the total firewood used in mezcal manufacturing.

\section{Conclusions}

Traditionally, mezcal is distilled over direct fire, which requires a high consumption of firewood. This process has several implications:

\textbf{Forest resource pressure}: Vegetation resource pressure and disruption of the main mechanism of plants' regeneration. Firewood overharvesting will reduce the shaded sites beneath perennial plants where most species (96\%) \cite{valiente} It takes approximately 10 kg of wood to produce a liter of mezcal.

\textbf{Health risks}: In some palenques, the lack of adequate ventilation can expose producers to smoke, compromising their health.

In this study, we compare two methods for distilling agave mosto: direct fire and electricity. We perform physical and chemical analyses and find no significant differences. Also, and more importantly, we perform double-anonymised sensory tests, in which we find that, statistically, traditional mezcal producers could not distinguish between fire- and electric-distilled spirits. The electrification of distillation offers a sustainable alternative that can mitigate the previously mentioned problems.

\textbf{Reducing dependence on firewood}: Electrification eliminates the need for firewood for distillation, reducing pressure on forests and greenhouse gas emissions.

\textbf{Improved working conditions}: Electrical distillation eliminates smoke and excessive heat from the traditional process, creating a safer and healthier working environment for producers.

In addition to distillation, electrification can improve other aspects of the production of mezcal:

\textbf{Temperature control during fermentation}: Solar panels can provide energy to heat the water needed for fermentation, especially in cold climates.

\textbf{Cooling systems}: Solar heaters can be used to cool the water used in the condensation process during distillation.

We emphasise that our research has shown that electrical distillation does not adversely affect spirit quality. A comparison between traditionally distilled mezcal and electrically distilled agave ferment found no significant differences in the sensory profile.
  
The electrification of the mezcal production presents an opportunity to combine tradition with technological innovation, promoting environmental sustainability, economic development, and social well-being in the producing communities.

Finally, we strongly suggest revising the Mexican Standard for the origin appeal of mezcal production to avoid using ''direct fire'' as an intrinsic characteristic of mezcal categories.

\subsection*{Author contributions}
.

Conceptualization: JAdRP; 

Investigation: JAdRP, ABO, ALO, PJVP, FdRP, MLS;

Formal Analysis: ABO, ALO, PJVP, FdRP, MLS;

Methodology: JAdRP, ALO, NYLM, PJVP, FdRP;

Resources; JAdRP, JAT, ALO, FdRP, MLS;

Project Administration; AVB, JAdRP;

Writing- Original Draft Preparation; JAdRP, ABO, ALO, PJVP;

Writing - Review \& Editing; FdRP, NYLM, AVB

\subsection*{Acknowledges}
 
CONAHCyT supported this work through the project number 319061. A. Balbuena Ortega acknowledges support from ``Investigadores por México'' SECIHTI.






\bibliography{reference}

@article{manuel88,
    author = {Manuel Martínez},
    title = {On the electrification of rural Mexico},
    journal = {Solar \& Wind Technology},
    year = {1988},
    volume = {5},
    number = {2},
    pages ={177-180}
}

@article{moreira,
    author = {D.F. Gómez-Hernández and B. Domenech and J. Moreir and N. Farrera and A. López-González and L. Ferrer-Martí},
    title = {Comparative evaluation of rural electrification project plans: A case study in Mexico},
    journal = {Energy Policy},
    year = {2019},
    volume = {129},
    pages ={23-33}
}

@article{BARAJASRAMIREZ2024,
title = {Influence of taste sensitivity on preference and sensory perception of mezcal},
journal = {Food Research International},
volume = {181},
pages = {114125},
year = {2024},
issn = {0963-9969},
doi = {https://doi.org/10.1016/j.foodres.2024.114125},
url = {https://www.sciencedirect.com/science/article/pii/S0963996924001959},
author = {J.A. Barajas-Ramírez and J. Pardo-Nuñez and V.G. Aguilar‐Raymundo and A.L. Gutiérrez-Salomón},
}

@book{olive,
    author = {L. Olive}, 
    title = {La ciencia y la tecnología en la sociedad del conocimiento. Ética, política y
epostemología},
    year = 2007,
    publisher = {Fondo de Cultura Económica},
    country = Mexico
}

@article{alcibar,
    author = {M. Alcíbar},
    title = {Comunicación pública de la ciencia y la tecnología: una aproximación crítica a su historia conceptual},
    journal = {Arbor},
    pages = {a242},
    volume = 191,
    number = 773,
    year = 2015,
    doi = {http://dx.doi.org/10.3989/arbor.2015.773n3012}
}

@misc{tradingeconomy,
    title = {Mexico Exports of beverages, spirits and vinegar},
    author = {Trading Economics},
    url = {https://tradingeconomics.com/mexico/exports/beverages-spirits-vinegar},
    date ={November, 2024},
    year = {2024}
}

@article{Schulz2007,
   author = {Hartwig Schulz and Malgorzata Baranska},
   doi = {10.1016/J.VIBSPEC.2006.06.001},
   issn = {0924-2031},
   issue = {1},
   journal = {Vibrational Spectroscopy},
   month = {1},
   pages = {13-25},
   publisher = {Elsevier},
   title = {Identification and quantification of valuable plant substances by IR and Raman spectroscopy},
   volume = {43},
   year = {2007},
}

@article{Lopez-rosas,
   author = {Agustín E. López-Rosas and Camila S. Gómez-Navarro and Walter M. Warren-Vega and Ana I. Zárate-Guzmán and Luis A. Romero-Cano},
   doi = {10.1016/J.JFCA.2024.106224},
   issn = {0889-1575},
   journal = {Journal of Food Composition and Analysis},
   month = {7},
   pages = {106224},
   publisher = {Academic Press},
   title = {Advancements towards the development artificial intelligence for sensory analysis: Integrating pattern recognition and signal processing in ATR-FTIR analysis of spirits},
   volume = {131},
   year = {2024},
}

@incollection{valiente,
    author ={Alfonso Valiente-Banuet},
    booktitle = {Mexican Fauna in the Anthropocene},
    publisher = {Springer} ,
    year = {2023},
title = {Mezcal boom and extinction debts},
    editor = {R Jones and C Ornelas-García and R Pineda-López and F Álvarez},
    isbn = {978-3-031-17277-9},
    doi = {https://doi.org/10.1007/978-3-031-17277-9}
}

@article{Seki2020,
   author = {Takakazu Seki and Kuo Yang Chiang and Chun Chieh Yu and Xiaoqing Yu and Masanari Okuno and Johannes Hunger and Yuki Nagata and Mischa Bonn},
   doi = {10.1021/ACS.JPCLETT.0C01259/ASSET/IMAGES/LARGE/JZ0C01259_0008.JPEG},
   issn = {19487185},
   issue = {19},
   journal = {Journal of Physical Chemistry Letters},
   month = {10},
   pages = {8459-8469},
   pmid = {32931284},
   publisher = {American Chemical Society},
   title = {The bending mode of water: A powerful probe for hydrogen bond structure of aqueous systems},
   volume = {11},
   url = {https://pubs.acs.org/doi/full/10.1021/acs.jpclett.0c01259},
   year = {2020}
}
\bibliographystyle{plain} 
\end{document}